\documentclass[conference]{IEEEtran}
\usepackage{listings}
\usepackage{url}

\usepackage[edges]{forest}

\usepackage{color}

\begin{document}
%
\title{On the Effectiveness of Clone Detection for Detecting IoT-related Vulnerable Clones}

\author{
Kentaro Ohno, 
Norihiro Yoshida, 
\IEEEauthorblockN{Wenqing Zhu 
and Hiroaki Takada}
\IEEEauthorblockA{Nagoya University, Japan\\}
\IEEEauthorblockA{\{k\_ohno, yoshida, zhuwqing1995, hiro\}@ertl.jp\\}
}

%


\maketitle

\begin{abstract}
Since IoT systems provide services over the Internet, they must continue to operate safely even if malicious users attack them.
Since the computational resources of edge devices connected to the IoT are limited, lightweight platforms and network protocols are often used.
Lightweight platforms and network protocols are less resistant to attacks, increasing the risk that developers will embed vulnerabilities.
The code clone research community has been developing approaches to fix buggy (e.g., vulnerable) clones simultaneously. However, there has been little research on IoT-related vulnerable clones. It is unclear whether existing code clone detection techniques can perform simultaneous fixes of the vulnerable clones.
In this study, we first created two datasets of IoT-related vulnerable code.  
We then conducted a preliminary investigation to show whether existing code clone detection tools  (e.g., NiCaD, CCFinderSW) are capable of detecting IoT-related vulnerable clones by applying them to the created datasets.
The preliminary result shows that the existing tools can detect them partially.
The datasets are available at: \url{https://zenodo.org/record/5090430#.YOqP_Oj7Q2w}.
\end{abstract}


%
\IEEEpeerreviewmaketitle

\section{Introduction}

\par
\par
\par
\par
\par


Internet of Things (IoT) envisions a self-configuring, adaptive, complex network that interconnects `things' to the Internet through the use of standard communication protocols \cite{minerva2015towards,Makhshari}.
The interconnected things have physical or virtual representation in the digital world, 
sensing/actuation capability, a programmability feature, and are uniquely identifiable \cite{minerva2015towards,Makhshari}. By 2020, Gartner estimates IoT connected devices will outnumber humans 4-to-1 \cite{hung2017leading}, and it is estimated that there will be more than 75 billion IoT connected devices worldwide by 2025 \cite{Makhshari,Statista}.
Since IoT systems provide services over the Internet, they must continue to operate safely even if malicious users attack them.
Since the computational resources of edge devices connected to the IoT are limited, lightweight platforms and network protocols are often used.
Lightweight platforms and network protocols are less resistant to attacks, increasing the risk that developers will embed vulnerabilities. Driven by the above considerations, the concept of IoT-related vulnerability is more complicated than traditional software systems \cite{Makhshari}. 
\par
So far, the software engineering and security research communities have been developing approaches to fix vulnerable clones simultaneously \cite{pham2010detecting, vuddy, Yoshida2010}. For example, Pham et al. developed SecureSync \cite{pham2010detecting}. This supporting tool is able to automatically detect recurring software vulnerabilities in different systems that share source code or libraries, which are the most frequent types of recurring vulnerabilities. Kim and Lee proposed VUDDY \cite{vuddy}, an approach to scalable and accurate code clone detection, which adopts robust parsing and a novel fingerprinting mechanism for functions. Despite the IoT-related vulnerabilities being more complex than other vulnerabilities described in the previous paragraph, little research has been done on IoT-related vulnerable clones. It is unclear whether existing code clone detection techniques can be used to perform simultaneous fixes of IoT-related vulnerable clones.
\par
In this study, we first present CVE-Based Dataset (\textbf{CBD}) and Issue-based Dataset (\textbf{IBD}) as datasets of IoT-related vulnerable code\footnotemark[1]. 
\textbf{CBD} is created by collecting programs obtained by searching CVE (Common Vulnerabilities and Exposures) IDs \cite{cve} of IoT-related vulnerabilities using the GitHub search engine. \textbf{IBD} is created by collecting files from pull requests and issues on IoT published by Makhshari and Mesbah \cite{Makhshari} before any patches were applied. 
We then conducted a preliminary investigation to show whether existing multilingual code clone detection tools (e.g., \textbf{NiCad} \cite{nicad},  \textbf{CCFinderSW} \cite{Semura2017,Semura2018,CCFinderSWURL}) are capable of detecting IoT-related vulnerable clones by applying them to the created datasets.
\par
Section \ref{2} describes how to create the proposed datasets, followed by an overview of the datasets in Section \ref{3}. Section \ref{4} shows the results of applying two code clone detection tools to the datasets as an example of their use. Section \ref{5} describes related works, and Section \ref{6} summarizes and discusses future work.

\section{Data Collection Methodology}\label{2}
The primary data of our datasets is the source code related to vulnerabilities. We begin by collecting source code reported as a vulnerability in GitHub by issues, comments, or pull requests. Then, we obtain the source code before vulnerability fixing using the Blame function of GitHub or the patch files.  According to the search approaches, we built two datasets: 
\begin{itemize}
    \item \textbf{CBD} (CVE-Based Dataset) 
    \item \textbf{IBD} (Issue-Based Dataset) 
\end{itemize}
The approaches are as follows.


\subsection{CVE-Based approach} \label{2-A}
\begin{figure}[tb]
  \centering
  \includegraphics[width=8cm]{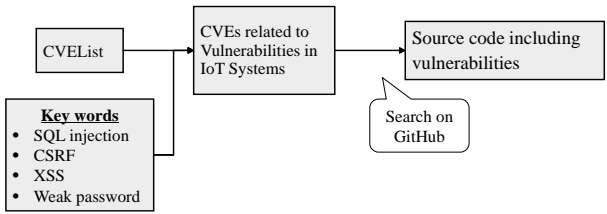}
  \caption{An overview of the process of creating \textbf{CBD} }
  \label{CBD_flow}
\end{figure}

This approach is based on CVEs (Common Vulnerabilities and Exposures) \cite{cve}.
Fig. \ref{CBD_flow} shows the process of building CBD. 
\subsubsection{\textbf{Search for the target keywords in the CVEList repository}} 
CVE is a list of records for publicly known cybersecurity and can be searched in CVEList \cite{cvelist}, built by CVE Numbering Authorities (CNAs).
From \cite{iiot}, there are three principal attack approaches for IoT systems: attacks via web management systems, attacks during remote software updates, and Mirai Attack.
Among them, attacks via web management systems appear most frequently.
Consequently, we selected the top four keywords reported in the attacks via the web management system.
: SQL injection, CSRF (Cross-Site Request Forgeries), XSS (Cross-Site Scripting), and weak passwords.
\par
\subsubsection{\textbf{Search for CVE-IDs that include the target keywords on GitHub}}
In this way, we collected the source code related to CVEs, including target keywords.
We checked the code, commit messages, issues, and pull requests of the search result to find the source code related to the target keywords \cite{githubsearch}.

\subsubsection{\textbf{Get the source code containing vulnerabilities}}
We collected source code before modification using the Blame function of GitHub or patch files.
\par
As a result of the search, there are CVEs for which the source code was unable to be obtained.
This may be because there are no changes based on that CVEs, or neither the source code nor the patch files with the vulnerability have not been uploaded to GitHub.

\subsection{Issue-Based approach}\label{2-B}
\begin{figure}[tb]
  \centering
  \includegraphics[width=8cm]{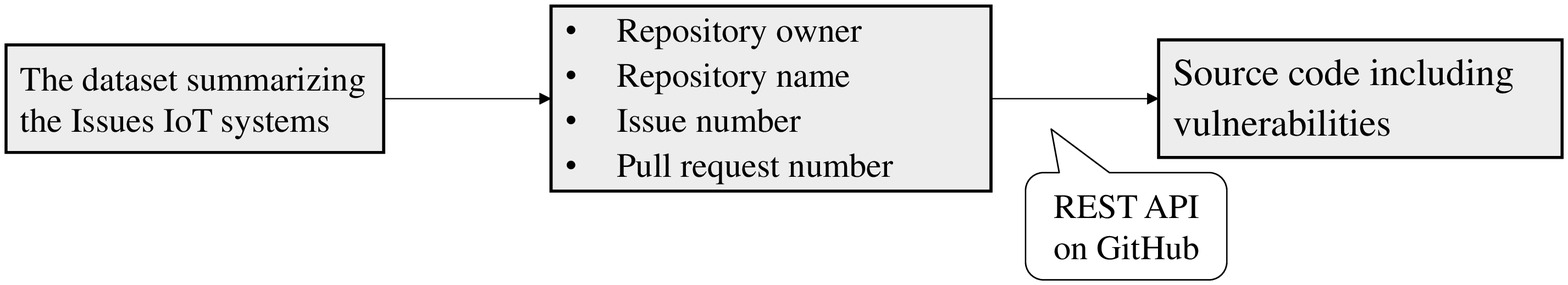}
  \caption{An overview of the process of creating \textbf{IBD} }
  \label{IBD_flow}
\end{figure}

This approach is based on bugs described in issues on GitHub. Fig. \ref{IBD_flow} shows the process of building IBD.
\subsubsection{\textbf{Get the data from the dataset of Makhshari and Mesbah}}
Get the repository owner, repository name, and issue number or pull request number from the dataset of Makhshari and Mesbah \cite{datasetofMakhshari}.
The dataset summarizes 5565 issues or pull requests related to IoT bugs.
Makhshari and Mesbah used this dataset to classify bugs in IoT systems.

\subsubsection{\textbf{Search for pull request number by issue number}}
We searched for pull request numbers by issue numbers using REST API on GitHub \cite{restapi}.
It is impossible to get the patch files or the files modified by the patch files directly from the issues.
Instead, we tried to search the pull request number 
by the repository owner, repository name, and issue number.
However, since the corresponding relationship between an issue and a pull request is not a  one-to-one correspondence, it is impossible to obtain all pull request numbers.


\textbf{curl -H ``Accept: application/vnd.github.cloak-preview'' [URL]\footnote{\textcolor{blue}{https://api.github.com/search/issues?q=repo:[owner]/[repo]/
+is:pr+is:closed+is:merged+comment:[issue number]}}}\\
With this command, it is able to get the pull request number.

\subsubsection{\textbf{Get the files modified by the patch file}}
From the acquired issue information and pull request number information, the patch files and the files changed by the patch files can be accessed by REST API on GitHub \cite{restapi}.

\textbf{curl -H ``Accept: application/vnd.github.v3+json'' [URL]\footnote{\textcolor{blue}{https://api.github.com/repos/[owner]/[repo]/pull/[pull request number]/files}}}\\
It is able to get the URL that points to the files modified by the patch files with this command.

\textbf{wget [URL]\footnote{\textcolor{blue}{https://github.com/[owner]/[repo]/pull/[pull request number].patch}}}\\
It is able to get the patch files with this command.

\textbf{wget [URL]\footnote{\textcolor{blue}{https://github.com/[owner]/[repo]/raw/[file number]/[file]}}}\\
It is able to get the files modified by the patch files with this command.
\par
\subsubsection{\textbf{Get the source code containing the vulnerability}}
We collected source code containing the vulnerability using patch files and the files changed by the patch files.

\textbf{patch -u  -R $<$ [patch file]} \\
It is able to get source code before modification with this command.

\section{Datasets}\label{3}
This section describes the size and composition of the CBD and IBD built in Section \ref{2}.
We searched 1800 CVEs by \ref{2-A} method, including 500 for SQL injection, 500 for CSRF, 500 for XSS, and 300 for weak passwords.
As a result, 20 CVEs are collected in \textbf{CBD}.
Table \ref{tab:CVEs} shows the list of the collected CVE-IDs.
\begin{table}[tb]
    \begin{center}
    \caption{ List of CVE-IDs in \textbf{CBD}}
    \begin{tabular}{|c|c|c|} \hline
    \textbf{SQL injection} & \textbf{XSS} & \textbf{CSRF} \\ \hline
    CVE-2010-1431 & CVE-2006-1230 & CVE-2012-1297 \\
    CVE-2014-3704 & CVE-2010-4513 & CVE-2012-1936 \\
    CVE-2014-7871 & CVE-2011-2711 & CVE-2015-1585 \\
    CVE-2014-8351 & CVE-2011-5081 & CVE-2020-8166 \\
    CVE-2015-2564 & CVE-2014-5464 & \\
    CVE-2015-4342 & CVE-2018-12588 & \\
    CVE-2015-7297 & CVE-2019-0213 & \\
    CVE-2015-8604 & CVE-2020-7996 & \\ \hline
        \end{tabular}
        \label{tab:CVEs}
    \end{center}
\end{table}

The number of Issues searched by \ref{2-B} approach is 5565 bugs.
The number of patch files in \textbf{IBD} collected by this approach is 619. 
Table \ref{tab:bugs} shows the number of patch files per repository.

\begin{table}[tb]

    \begin{center}
        \caption{The Number of files in each repository included in the IBD }
        \begin{tabular}{|l|l|l|} \hline
            \textbf{owner/repository} & \textbf{number of files} \\ \hline
            eclipse/kapua & 386 patch files \\
            eclipse/hono & 101 patch files \\
            AlCalzone/node-tradfri-client & 47 patch files \\
            GladysAssistant/Gladys & 24 patch files \\
            mainflux/mainflux & 18 patch files \\
            kubeedge/kubeedge & 10 patch files \\
            eclipse/kura & 7 patch files \\
            particle-iot/device-os & 7 patch files \\
            yodaos-project/yoda.js & 4 patch files \\
            openhab/openhab-core & 3 patch files \\
            ikalchev/HAP-python & 2 patch files \\
            OpenZWave/Zwave2Mqtt & 2 patch files \\
            Azure/azure-iot-sdk-c & 1 patch file \\
            codetheweb/tuyapi & 1 patch file \\
            eclipse/upm & 1 patch file \\
            homieiot/homie-esp8266 & 1 patch file \\
            MCS-Lite/mcs-lite-app & 1 patch file \\
            mysensors/NodeManager & 1 patch file \\
            openhab/openhab-ios & 1 patch file \\ 
            pihome-shc/pihome & 1 patch file \\ \hline
        \end{tabular}
        \label{tab:bugs}
    \end{center}

\end{table}

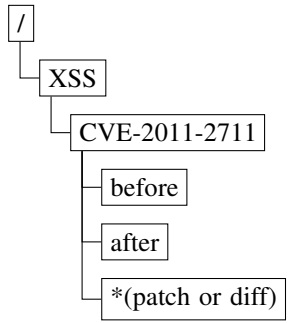
\begin{figure}[tb]
    \centering
    \begin{forest}
 for tree={grow'=0,folder,draw}
 [/
  [XSS
    [CVE-2011-2711
      [before]
      [after]
      [*(patch or diff)]
    ]
  ]
 ]
\end{forest}
    \caption{\textbf{CBD} directory hierarchy (excerpt of the part containing CVE-2011-2711)}
    \label{fig:CBDtree}
\end{figure}

\textbf{CBD} has 20 CVEs related to IoT systems.
In the XSS, SQL injection, and CSRF directories, files before and after the modifying are included.
Fig. \ref{fig:CBDtree} shows a part of the structure of \textbf{CBD}.

\begin{figure}[!t]
    \centering
    \begin{forest}
 for tree={grow'=0,folder,draw}
 [/
  [1548
    [1548.patch]
    [AssetTabItem.java]
    [DeviceTabItem.java]
    [TopicsTabItem.java]
  ]
 ]
\end{forest}
    \caption{\textbf{IBD} directory hierarchy (excerpt of the part containing 1548)}
    \label{fig:IBDtree}
\end{figure}
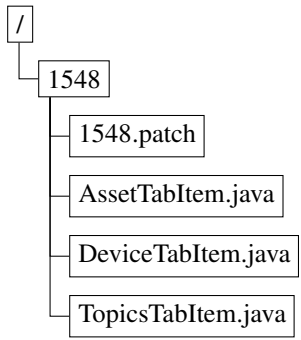

\textbf{IBD} has 619 patch files used to fix the vulnerability of all files.
A directory with the pull request number is created, and the patch file and the files modified by the patch file are included. 
Fig. \ref{fig:IBDtree} shows a part of the structure of \textbf{IBD}.

We have put the dataset into a JSON file. This file is included in \textbf{IBD}.
\par
The file contains the following fields:
\begin{itemize}
    \item ``ID''
    \item ``owner'' : repository owner
    \item ``repo'' : repository name
    \item ``number'' : ID of a pull request 
    \item ``locate'' : directory name
    \item ``title'' : The title of the pull request
    \item ``refer'' : The URL of the pull request
    \item ``file'' : The files containing vulnerabilities
\end{itemize}

\section{Preliminary Investigation}\label{4}
\begin{figure}[b]
\centering
\includegraphics[clip, width=3.4in]{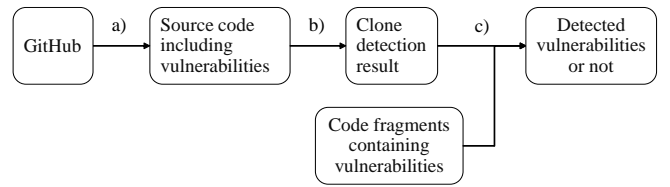}
\caption{An overview of vulnerabilities detection using code clone detection tools}
\label{tab:detectflow}
\end{figure}

Based on \textbf{CBD} and \textbf{IBD}, we investigated whether vulnerabilities in IoT systems could be detected using code clone detection tools.
Fig. \ref{tab:detectflow} shows the process of vulnerabilities detection by code clone detection.
This process is mainly comprised of the following three steps:
\renewcommand{\theenumi}{\alph{enumi}}
\begin{enumerate}
    \item \textbf{Datasets creation} \\
    See Section \ref{2} and Section \ref{3}.
    \item \textbf{Clone detection} \\
    We used NiCad \cite{nicad} and CCFinderSW \cite{Semura2017,Semura2018,CCFinderSWURL} to detect code clones in source code from \textbf{CBD} and \textbf{IBD}.
    \item \textbf{Vulnerability judgment}  \\
    We compared the source code reported as code clones with the vulnerabilities contained in the patch files to judge whether the vulnerabilities have been detected.
\end{enumerate}

\subsection{Settings}
\subsubsection{Investigation target}
CVEs with multiple files in \textbf{CBD} (8 CVE-IDs in total) and patch files with multiple modifications in \textbf{IBD} (randomly choose 12 patch files) were investigated.
If only one code fragment contains a vulnerability, the code fragment containing the vulnerability is unable to be detected.
In \textbf{IBD}, the entire source code was investigated by following the procedure below to conduct an investigation, assuming that detection would be performed.
\begin{itemize}
    \item \textbf{Select patch files: }
We selected the target patch files randomly.
    \item \textbf{Git clone: }
Copy the entire cloned software to a directory.
\textbf{IBD} contains the patch files and the unmodified files. Therefore, the overlapping files are skipped.
\end{itemize}


\subsubsection{Criterion}
\begin{figure}[tb]
\centering
\includegraphics[clip, width=3.4in]{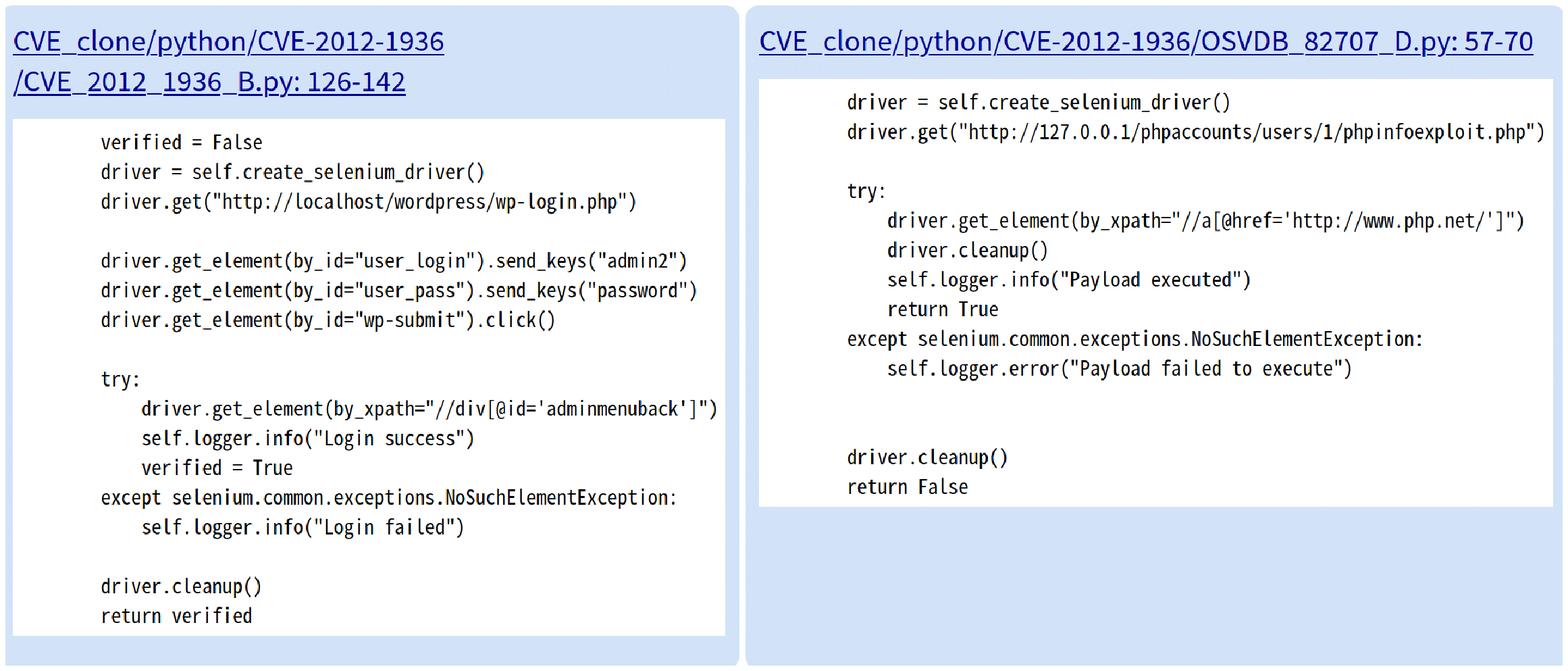}
\caption{An example of the clone detection result on \textbf{NiCad}}
\label{fig:3.4}
\end{figure}

If the detected code clone matches or resembles a code fragment containing the vulnerability, it is judged that the vulnerability has been detected.
Fig. \ref{fig:3.4} shows an example of the detection result on \textbf{NiCad}, and Listing \ref{diff} shows an example of a diff file.
In such a case, it is judged that the vulnerability could be detected because the detection result of the code clone matches the modified code fragment.

\begin{figure}
    \centering
    \begin{lstlisting}[caption=CVE-2012-1936.diff,label=diff]
#include<stdio.h>
--- before/CVE_2012_1936_B.py	2020-11-02 23:34:53.347981719 -0800
+++ after/CVE_2012_1936_B.py	2020-11-02 23:34:53.347981719 -0800
@@ -124,7 +124,7 @@
 
     def verify(self):
         verified = False
-        driver = self.create_selenium_driver()
+        driver = self.create_selenium_driver(javascript=False)
         driver.get("http://localhost/wordpress/wp-login.php")
 
         driver.get_element(by_id="user_login").send_keys("admin2")
\end{lstlisting}
\end{figure}

\subsubsection{Clone detection tools}
We chose \textbf{NiCad} and \textbf{CCFinderSW} in this investigation because of the need for multiple languages support.
\textbf{NiCad} detects code clones by functions or blocks and can detect Type-1, Type-2 and Type-3 \cite{ROY2009470} code clones.
As a token-based detection tool, \textbf{CCFinderSW} can detect Type-1 and Type-2 code clones by matching token sequences consecutively.
\par
We configured the two tools as shown in Table \ref{tab:nicad_setting} and Table \ref{tab:ccfinder_setting}.

\begin{table}[tb]
    \begin{center}
    \caption{Settings of \textbf{NiCad}}
  \begin{tabular}{rl}\hline
    granularity: & Blocks or Functions \\
    similarity: & 40\% or 70\% \\
    Type: & Type-3  \\
    \hline
  \end{tabular}
  \label{tab:nicad_setting}
  \end{center}
\end{table}

\begin{table}[tb]
    \begin{center}
    \caption{Settings of \textbf{CCFinderSW}}
  \begin{tabular}{rl}\hline
    granularity: & Tokens \\
    Minimum number of tokens: & 50 \\
    Type: & Type-2 \\
    \hline
  \end{tabular}
  \label{tab:ccfinder_setting}
  \end{center}
\end{table}

\subsection{Investigation Using \textbf{CBD}}
In \textbf{CBD}, clone detection was performed for each CVE file group containing vulnerabilities using \textbf{NiCad} and \textbf{CCFinderSW}. In \textbf{NiCad}, we compared when the similarity was 40\% and 70\%.
In the evaluation of \textbf{CBD}, the items that serve as investigation indicators are defined as shown in Table \ref{tab:dataset1}.
\begin{table}[tb]
    \begin{center}
        \caption{Indicators for the investigation using \textbf{CBD}}
        \begin{tabular}{ll} \hline
            A: & Number of files \\ 
            B: & Number of files in the same clone group \\ 
               & as the code fragment containing the vulnerability \\
            C: & Number of clones containing vulnerabilities in B \\
            D: & C divide by B \\
            E: & C divide by A \\ \hline
        \end{tabular}
        \label{tab:dataset1}
    \end{center}
\end{table}
The indicator D shows the percentage of code clones containing vulnerabilities in the
detection results. 
The indicator E shows the percentage of code clones that contain the vulnerabilities found in the files. 

\subsection{Investigation Using \textbf{IBD}}
In \textbf{IBD}, we detect the vulnerabilities using \textbf{NiCad} (similarity threshold is set to 70\%). The granularities are blocks and functions.
In the evaluation of \textbf{IBD}, the items that serve as investigation indicators are defined as shown in Table \ref{tab:dataset2}.

\begin{table}[tb]
    \begin{center}
        \caption{Indicators for the investigation using \textbf{IBD}}
        \begin{tabular}{ll} \hline
            A: & Number of vulnerable code in patch file \\
            B: & Number of code fragment in the same clone group \\ 
               & as the code fragment containing the vulnerability \\
            C: & Number of clones containing vulnerabilities in B \\ 
            D: & C divide by B \\
            E: & C divide by A \\   \hline
        \end{tabular}
        \label{tab:dataset2}
    \end{center}
\end{table}
The indicator D means precision.
This shows the percentage of code clones containing vulnerabilities in the detection results.
The indicator E means recall.
This shows the percentage of code clones that contain the vulnerabilities found in the patch.

\subsection{Results}
\subsubsection{Result from \textbf{CBD}}
The evaluation results in \textbf{NiCad} are shown in Table \ref{tab:dataset1nicad40} and Table \ref{tab:dataset1nicad70}, and the evaluation results in \textbf{CCFinderSW} are shown in Table \ref{tab:dataset1ccfindersw}.

\textbf{NiCad} and \textbf{CCFinderSW} were able to detect 84\% of CVE-2011-2711. \textbf{NiCad} (similarity threshold is set to 40\%) was able to detect all CVE-2012-1936, but \textbf{NiCad} (similarity threshold is set to 70\%) was unable to. \textbf{CCFinderSW} was able to detect 50\% of CVE-2012-1936. \textbf{NiCad} was unable to detect CVE-2015-8604 and CVE-2020-7996, but \textbf{CCFindeSW} could. Neither \textbf{NiCad} nor \textbf{CCFindeSW} could detect CVE-2014-8351, CVE-2015-4342, CVE-2019-0213, and CVE-2020-8166.

\begin{table}[tb]
  \begin{center}
    \caption{Evaluation result in \textbf{NiCad} (40\%) for \textbf{CBD}}
    \begin{tabular}{|l|r|r|r|r|r|} \hline
      CVE-ID & (A) & (B) & (C) &  (D) & (E)
 \\ \hline \hline
      CVE-2011-2711 (C) & 25 & 25 & 21 & 0.840 & 0.840 \\
      CVE-2012-1936 (Python) & 2 & 3 & 2 & 0.667 & 1.000 \\
      CVE-2015-8604 (PHP) & 2 & 0 & 0 &  & 0 \\
      CVE-2020-7996 (PHP) & 2 & 0 & 0 &  & 0 \\
      CVE-2014-8351 (PHP) & 2 & 0 & 0 &  & 0 \\
      CVE-2015-4342 (PHP) & 2 & 0 & 0 &  & 0 \\
      CVE-2019-0213 (Java) & 4 & 0 & 0 &  & 0 \\
      CVE-2020-8166 (Ruby) & 2 & 0 & 0 &  & 0 \\ \hline
    \end{tabular}
    \label{tab:dataset1nicad40}
  \end{center}
\end{table}

\begin{table}[tb]
  \begin{center}
    \caption{Evaluation result in \textbf{NiCad} (70\%) for \textbf{CBD}}
    \begin{tabular}{|l|r|r|r|r|r|} \hline
      CVE-ID & (A) & (B) & (C) &  (D) & (E)
 \\ \hline \hline
      CVE-2011-2711 (C) & 25 & 25 & 21 & 0.840 & 0.840 \\
      CVE-2012-1936 (Python) & 2 & 0 & 0 & 0 & 0 \\
      CVE-2015-8604 (PHP) & 2 & 0 & 0 &  & 0 \\
      CVE-2020-7996 (PHP) & 2 & 0 & 0 &  & 0 \\
      CVE-2014-8351 (PHP) & 2 & 0 & 0 &  & 0 \\
      CVE-2015-4342 (PHP) & 2 & 0 & 0 &  & 0 \\
      CVE-2019-0213 (Java) & 4 & 0 & 0 &  & 0 \\
      CVE-2020-8166 (Ruby) & 2 & 0 & 0 &  & 0 \\ \hline
    \end{tabular}
    \label{tab:dataset1nicad70}
  \end{center}
\end{table}

\begin{table}[tb]
  \begin{center}
    \caption{Evaluation result in \textbf{CCFinderSW} for \textbf{CBD}}
    \begin{tabular}{|l|r|r|r|r|r|} \hline
      CVE-ID & (A) & (B) & (C) &  (D) & (E)
 \\ \hline \hline
      CVE-2011-2711 (C) & 25 & 21 & 21 & 1.000 & 0.840 \\
      CVE-2012-1936 (Python) & 2 & 2 & 1 & 0.500 & 0.500 \\
      CVE-2015-8604 (PHP) & 2 & 2 & 2 & 1.000 & 1.000 \\ 
      CVE-2020-7996 (PHP) & 2 & 2 & 2 & 1.000 & 1.000 \\
      CVE-2014-8351 (PHP) & 2 & 0 & 0 &  & 0 \\
      CVE-2015-4342 (PHP) & 2 & 0 & 0 &  & 0 \\
      CVE-2019-0213 (Java) & 4 & 0 & 0 &  & 0 \\
      CVE-2020-8166 (Ruby) & 2 & 0 & 0 &  & 0 \\ \hline
    \end{tabular}
    \label{tab:dataset1ccfindersw}
  \end{center}
\end{table}

\subsubsection{Result from \textbf{IBD}}
The investigation results in \textbf{IBD} are shown in Table \ref{tab:dataset2nicadfunction} and \ref{tab:dataset2nicadblock}.
\textbf{NiCad} was able to detect an average of 32\% of vulnerabilities by function and block. On the other hand, precision was 0.85 by function detection and 0.77 by block detection, showing a difference.

\begin{table}[tb]
  \begin{center}
    \caption{Evaluation result in \textbf{NiCad} (functions) for \textbf{IBD}}
    \begin{tabular}{|c|r|r|r|r|r|r|r|} \hline
      patch & (A) & (B) & (C) &  (D) & (E)

 \\ \hline \hline
      682 & 6 & 0 & 0 &  & 0.00 \\
      712 & 4 & 6 & 4 & 0.67 & 1.00 \\
      1223 & 11 & 2 & 2 & 1.00 & 0.18 \\
      1324 & 8 & 22 & 6 & 0.27 & 0.75 \\
      1548 & 6 & 6 & 6 & 1.00 & 1.00 \\
      1868 & 5 & 2 & 2 & 1.00 & 0.40 \\
      2065 & 2 & 0 & 0 &  & 0.00 \\
      2240 & 9 & 2 & 2 & 1.00 & 0.22 \\
      2351 & 3 & 0 & 0 &  & 0.00 \\
      2364 & 2 & 0 & 0 &  & 0.00 \\
      2531 & 2 & 0 & 0 &  & 0.00 \\
      2834 & 8 & 2 & 2 & 1.00 & 0.25 \\ \hline
      average & 5.5 & 3.5 & 2.0 & 0.85 & 0.32 \\ \hline
    \end{tabular}
    \label{tab:dataset2nicadfunction}
  \end{center}
\end{table}

\begin{table}[tb]
  \begin{center}
    \caption{Evaluation result in \textbf{NiCad} (blocks) for \textbf{IBD}}
    \begin{tabular}{|c|r|r|r|r|r|r|r|} \hline
      patch & (A) & (B) & (C) &  (D) & (E)

 \\ \hline \hline
      682 & 6 & 0 & 0 &  & 0.00 \\
      712 & 4 & 8 & 4 & 0.50 & 1.00 \\
      1223 & 11 & 2 & 2 & 1.00 & 0.18 \\
      1324 & 8 & 24 & 6 & 0.25 & 0.75 \\
      1548 & 6 & 6 & 6 & 1.00 & 1.00 \\
      1868 & 5 & 2 & 2 & 1.00 & 0.40 \\
      2065 & 2 & 0 & 0 &  & 0.00 \\
      2240 & 9 & 3 & 2 & 0.67 & 0.22 \\
      2351 & 3 & 0 & 0 &  & 0.00 \\
      2364 & 2 & 0 & 0 &  & 0.00 \\
      2531 & 2 & 0 & 0 &  & 0.00 \\
      2834 & 8 & 2 & 2 & 1.00 & 0.25 \\ \hline
      average & 5.5 & 3.92 & 2.0 & 0.77 & 0.32 \\ \hline
    \end{tabular}
    \label{tab:dataset2nicadblock}
  \end{center}
\end{table}

\par
Some vulnerabilities were able to be detected, but some were unable to be detected.
This investigation shows that clone detection tools can partially detect vulnerabilities in IoT systems.
\subsection{Discussion}
As a result of this investigation, it was found that vulnerabilities in IoT systems can be partially detected.
This section will discuss the situations for vulnerabilities not detected in two aspects: datasets and detection tools.

\subsubsection{Datasets}
We visually considered the cause of the vulnerabilities in the IoT system that was unable to be detected.
The first cause is that there may be only one code fragment containing the vulnerability in the project.
Because there is no other code fragment contains the same vulnerability, there would be no code clone result.
\par
The second cause is that it is unable to be detected due to the setting of the clone detection tools.
Clone detection tools can often set the degree of similarity by parameterizing how similar the code fragments are compared.
The lower the similarity threshold is, the more vulnerable code clones could be detected, but the more code fragments will be reported as clones falsely.
Based on the recommended configurations, we set the similarity threshold of \textbf{Nicad} to 70\% and set the minimum number of matching tokens of \textbf{CCFinderSW} to 50. 
If the code fragments containing vulnerabilities are smaller than 50 tokens, or the similarity is lower than 70\%, they are unable to be detected by the tools.
In this case, we need to consider the balance between precision and recall.
\par
The third cause is that a vulnerability in a project may consist of multiple files. 
In such a case, it is necessary to detect code clones consists of several code fragments from multiple files.
Because the used tools are unable to detect clones consist of multiple files, this part is unable to be detected.

\subsubsection{Detection Tools}
Different detection tools have different features. It is necessary to find out which kind of detection tools are better for detecting vulnerabilities.
In this investigation, we used two clone detection tools, \textbf{NiCad} and \textbf{CCFinderSW}.
\textbf{NiCad} detects code clones by functions or blocks and can detect Type-1, Type-2, and Type-3 code clones.
As a token-based detection tool, \textbf{CCFinderSW} can detect Type-1 and Type-2 code clones.
\par

Comparing the results of the two clone detection tools in \textbf{CBD}, detection tools with token sequence granularity and detection tools with Type-3 detection ability show effectiveness in detecting vulnerabilities in IoT systems.
For example, CVE-2015-8604 and CVE-2020-7996 are only detected by \textbf{CCFinderSW}, the detection tool with token sequences granularity.
They were unable to be detected in functions granularity because CVE-2015-8604 does not have a symbol indicating function, and CVE-2020-7996 has only a few lines of vulnerability code for one function. 
As for the detection ability of Type-3 clones, when detection clones in CVE-2012-1936, \textbf{NiCad} detected all vulnerabilities, but \textbf{CCFinderSW} only detected half of them.
\par
Investigation of \textbf{IBD} shows that detection by functions is effective in detecting vulnerabilities in IoT systems.
This is because when clone detection was performed on 712, 1324, and 2240 in \textbf{IBD}, Precision in function granularity was higher than in block granularity. 
On the other hand, the detection at function granularity showed the possibility of detecting the same degree of vulnerabilities as the block granularity. 

\section{Related Work} \label{5}
\subsection{Clone detection tools}
Various clone detection tools have been developed to solve code clone problems such as defeats propagation in software projects.
According to source code normalization and similarity calculation, the detection ability of tools is different.
Typical features include the types of code clones that can be detected and granularity. 
In addition, many tools can set the degree of similarity by parameterizing how similar the code fragments are.
The lower the similarity threshold is, the more vulnerable code clones could be detected, but the more code fragments will be reported as clones falsely.
Furthermore, different tools may support different programming languages.
\textbf{CCFinderSW} \cite{Semura2017,Semura2018,CCFinderSWURL} supports various languages by giving a syntax definition description as input. \textbf{NiCad} \cite{nicad} (ver. 6.2) supports C, C\#, Java, Python, PHP, Ruby, WSDL, and ATL.
\textbf{CCFinderSW} can detect Type-1 and Type-2 \cite{ROY2009470} code clones, and \textbf{NiCad} can detect Type-1, Type-2 and Type-3 \cite{ROY2009470} code clones.


\label{cc}

\textbf{CCFinderSW} \cite{Semura2017} \cite{Semura2018} is a clone detection tool that can detect code clones in various programming languages.
This is achieved by giving the syntax definition description of ANTLR, one of the parser generator generation systems, as an input.
\textbf{CCFinderSW} performs clone detection through three steps: lexical analysis, variable processing, and clone detection/output shaping. 
\par


\noindent
\par
\par
\textbf{NiCad} \cite{nicad} is a clone detection tool that uses a hybrid approach to generate structurally meaningful Type-3 clones by combining language-dependent analysis and language-independent similarity analysis.
\textbf{NiCad} has three main stages: analysis, normalization, and comparison/cloning detection. 
\par
\par
\par

\subsection{VUDDY}
VUDDY \cite{vuddy} is a tool for the scalable detection of vulnerable code clones in C and C++, which is capable of detecting security vulnerabilities in large software programs efficiently and accurately.
VUDDY searches for vulnerabilities by comparing functions from a dictionary of vulnerability with target files and detecting clones.
Clone detection tools aim to detect code clones, while vulnerability detection tools, including VUDDY, aim to find vulnerabilities. Therefore, the detection results are also different.
The clone detection tool detects $N$($N\geq0$) of similar code clones containing $M$($M\geq2$) of code fragments. On the other hand, VUDDY detects code fragments with $M$($M\geq0$) vulnerabilities for one of the $N$($N\geq0$) functions in the vulnerability code dictionary.
\par
VUDDY detects vulnerabilities through two stages: preprocessing and clone detection. 
The preprocessing consists of functions search, abstraction/standardization, and fingerprint generation and must be performed on functions of target files and the vulnerability code dictionary. However, if the vulnerability code dictionary is not added, the functions of the vulnerability code dictionary can be reused by preprocessing as a dictionary once.
Clone detection consists of key search and hash search. The key search searches for the same key in the dictionary, and if there is the same key, the hash search is performed. The Hash search searches for the same hash value in the same key, and if there is the same hash value, it is judged as a code clone.

\section{Summary} \label{6}
In this paper, we collected datasets of vulnerabilities in IoT systems. We used the datasets to confirm the effectiveness of the clone detection tool in detecting vulnerabilities in IoT systems. 
It turned out that the vulnerability can be partially detected.
In addition, we considered the reasons why the vulnerability was unable to be detected and the detection rate due to the characteristics of clone detection.

As a future work, we are considering developing a tool that can detect vulnerabilities in IoT systems using clone detection technology. Therefore, it is necessary to add datasets and examine the tool configuration to increase the precison and recall.


\section*{Acknowledgment}
We thank Mr. Ryu Miyaki and Mr. Junyu Chen at Nagoya University for their helpful comments on this paper.
This work was supported by JSPS KAKENHI Grant Numbers JP20K11745.



%



\begin{thebibliography}{1}
\bibitem[1]{minerva2015towards} R. Minerva, A. Biru, and D. Rotondi, “Towards a definition of the
Internet of Things (IoT),” IEEE Internet Initiative, vol. 1, pp. 1–86,
2015.
\bibitem[2]{Makhshari} A. Makhshari and A. Mesbah, “IoT bugs and development challenges,”
in 2021 IEEE/ACM 43rd International Conference on Software Engineering (ICSE). IEEE, 2021, pp. 460–472.
\bibitem[3]{hung2017leading} M. Hung, “Leading the iot, gartner insights on how to lead in a connected
world,” Gartner Research, vol. 1, pp. 1–5, 2017.
\bibitem[4]{Statista} Statista. Internet of Things (IoT) connected devices installed base worldwide from 2015 to 2025 (in billions). [Online]. Available:  https://www.statista.com/statistics/471264/iot-number-of-connected-devices-worldwide/
\bibitem[5]{pham2010detecting} N. H. Pham, T. T. Nguyen, H. A. Nguyen, X. Wang, A. T. Nguyen,
and T. N. Nguyen, “Detecting recurring and similar software vulnerabilities,” in 2010 ACM/IEEE 32nd International Conference on Software
Engineering, vol. 2. IEEE, 2010, pp. 227–230.
\bibitem[6]{vuddy} S. Kim and H. Lee, “Software systems at risk: An empirical study of
cloned vulnerabilities in practice,” Comput. Secur., vol. 77, pp. 720–736,
2018. [Online]. Available: \url{https://doi.org/10.1016/j.cose.2018.02.007}
\bibitem[7]{Yoshida2010} N. Yoshida, T. Hattori, and K. Inoue, “Finding Similar Defects
Using Synonymous Identifier Retrieval,” in Proceedings of the 4th
International Workshop on Software Clones, ser. IWSC ’10. New
York, NY, USA: Association for Computing Machinery, 2010, p.
49–56. [Online]. Available: \url{https://doi.org/10.1145/1808901.1808908}
\bibitem[8]{cve} MITRE, “CVE - CVE,” \url{https://cve.mitre.org/}, (Accessed on
07/04/2021).
\bibitem[9]{nicad} J. R. Cordy and C. K. Roy, “The NiCad clone detector,” in 2011 IEEE
19th International Conference on Program Comprehension. IEEE,
2011, pp. 219–220.
\bibitem[10]{Semura2017}Y. Semura, N. Yoshida, E. Choi, and K. Inoue, “CCFinderSW: Clone
Detection Tool with Flexible Multilingual Tokenization,” in 2017 24th
Asia-Pacific Software Engineering Conference (APSEC), 2017, pp. 654–
659.
\bibitem[11]{Semura2018} ——, “Multilingual Detection of Code Clones Using ANTLR Grammar
Definitions,” in 2018 25th Asia-Pacific Software Engineering Conference
(APSEC), 2018, pp. 673–677.
\bibitem[12]{CCFinderSWURL} Y. Semura, “CCFinderSW,” \url{https://github.com/YuichiSemura/}
CCFinderSW.
\bibitem[13]{cvelist} CVE Program, “CVEProject/cvelist: Pilot program for CVE submission
through GitHub,” \url{https://github.com/CVEProject/cvelist}, (Accessed on
07/04/2021).
\bibitem[14]{iiot} X. Jiang, M. Lora, and S. Chattopadhyay, “An experimental analysis of
security vulnerabilities in industrial IoT devices,” ACM Transactions on
Internet Technology (TOIT), vol. 20, no. 2, pp. 1–24, 2020.
\bibitem[15]{githubsearch} GitHub, “GitHub · Where software is built,” \url{https://github.com/search/advanced}, (Accessed on 07/04/2021).
\bibitem[16]{datasetofMakhshari} A. Makhshari and A. Mesbah, “IoTSEstudy/IoTbugschallenges: Replication Package of ”IoT Bugs and Development Challenges” study,” \url{https://github.com/IoTSEstudy/IoTbugschallenges}, (Accessed on 07/04/2021).
\bibitem[17]{restapi} GitHub, “GitHub REST API - GitHub Docs,” \url{https://docs.github.com/en/rest}, (Accessed on 07/04/2021).
\bibitem[18]{ROY2009470} C. K. Roy, J. R. Cordy, and R. Koschke, “Comparison and evaluation
of code clone detection techniques and tools: A qualitative approach,”
Science of Computer Programming, vol. 74, no. 7, pp. 470–495, 2009.
[Online]. Available: \url{https://www.sciencedirect.com/science/article/pii/S0167642309000367}
\end{thebibliography}

\end{document}